\documentclass[aps,prl,superscriptaddress,showpacs,floatfix,twocolumn]{revtex4}

\usepackage{graphicx}	
\usepackage{xspace}

\newcommand{\sNN}     {\mbox{$\sqrt{s_{_{\rm NN}}}$}\xspace}
\newcommand{\pT}      {\mbox{$p_{\rm T}$}\xspace}
\newcommand{\kT}      {\mbox{$k_{\rm T}$}\xspace}

\newcommand{\mT}      {\mbox{$m_{\rm T}$}\xspace}
\newcommand{\mmT}     {\mbox{$\langle m_{\rm T}\rangle$}\xspace}

\newcommand{\bfq}     {\mbox{$\mathbf{q}$}\xspace}
\newcommand{\Np}      {\mbox{$N_{\rm part}$}\xspace}
\newcommand{\Npc}     {\mbox{$N_{\rm part}^{1/3}$}\xspace}
\newcommand{\Rs}      {\mbox{$R_{\rm side}$}\xspace}
\newcommand{\Ro}      {\mbox{$R_{\rm out}$}\xspace}
\newcommand{\Rl}      {\mbox{$R_{\rm long}$}\xspace}
\newcommand{\RsS}     {\mbox{$R^{2}_{\rm side}$}\xspace}
\newcommand{\RoS}     {\mbox{$R^{2}_{\rm out}$}\xspace}
\newcommand{\RlS}     {\mbox{$R^{2}_{\rm long}$}\xspace}

\newcommand{\Ros}     {\mbox{$R_{\rm out}/R_{\rm side}$}\xspace}

\newcommand{\qi}      {\mbox{$q_{\rm inv}$}\xspace}
\newcommand{\qs}      {\mbox{$q_{\rm side}$}\xspace}
\newcommand{\qo}      {\mbox{$q_{\rm out}$}\xspace}
\newcommand{\ql}      {\mbox{$q_{\rm long}$}\xspace}
\newcommand{\qsS}     {\mbox{$q^{2}_{\rm side}$}\xspace}
\newcommand{\qoS}     {\mbox{$q^{2}_{\rm out}$}\xspace}
\newcommand{\qlS}     {\mbox{$q^{2}_{\rm long}$}\xspace}

\newcommand{\cisq}    {\mbox{$\chi^{\rm 2}$}\xspace}

\newcommand{\CqB}     {\mbox{$C_{2}(\bf q)$}\xspace}
\newcommand{\CqI}     {\mbox{$C_{2}(q_{\rm inv})$}\xspace}
\newcommand{\SrB}     {\mbox{$S(\bf r)$}\xspace}
\newcommand{\SrI}     {\mbox{$S(r)$}\xspace}

\newcommand{\KernB}   {\mbox{$K(\bf q,r)$}\xspace}

\begin{document}


\title{Kaon interferometric probes of space-time evolution \\
in Au+Au collisions at $\sNN$ = 200 GeV}

\newcommand{\abilene}{Abilene Christian University, Abilene, TX 79699, U.S.}
\newcommand{\banaras}{Department of Physics, Banaras Hindu University, Varanasi 221005, India}
\newcommand{\bnlphys}{Brookhaven National Laboratory, Upton, NY 11973-5000, U.S.}
\newcommand{\caucr}{University of California - Riverside, Riverside, CA 92521, U.S.}
\newcommand{\cns}{Center for Nuclear Study, Graduate School of Science, University of Tokyo, 7-3-1 Hongo, Bunkyo, Tokyo 113-0033, Japan}
\newcommand{\colorado}{University of Colorado, Boulder, CO 80309, U.S.}
\newcommand{\columbia}{Columbia University, New York, NY 10027 and Nevis Laboratories, Irvington, NY 10533, U.S.}
\newcommand{\dapnia}{Dapnia, CEA Saclay, F-91191, Gif-sur-Yvette, France}
\newcommand{\debrecen}{Debrecen University, H-4010 Debrecen, Egyetem t{\'e}r 1, Hungary}
\newcommand{\elte}{ELTE, E{\"o}tv{\"o}s Lor{\'a}nd University, H - 1117 Budapest, P{\'a}zm{\'a}ny P. s. 1/A, Hungary}
\newcommand{\fsu}{Florida State University, Tallahassee, FL 32306, U.S.}
\newcommand{\gsu}{Georgia State University, Atlanta, GA 30303, U.S.}
\newcommand{\hiroshima}{Hiroshima University, Kagamiyama, Higashi-Hiroshima 739-8526, Japan}
\newcommand{\ihepprot}{IHEP Protvino, State Research Center of Russian Federation, Institute for High Energy Physics, Protvino, 142281, Russia}
\newcommand{\illuiuc}{University of Illinois at Urbana-Champaign, Urbana, IL 61801, U.S.}
\newcommand{\isu}{Iowa State University, Ames, IA 50011, U.S.}
\newcommand{\jinrdubna}{Joint Institute for Nuclear Research, 141980 Dubna, Moscow Region, Russia}
\newcommand{\kaeri}{KAERI, Cyclotron Application Laboratory, Seoul, Korea}
\newcommand{\kek}{KEK, High Energy Accelerator Research Organization, Tsukuba, Ibaraki 305-0801, Japan}
\newcommand{\kfki}{KFKI Research Institute for Particle and Nuclear Physics of the Hungarian Academy of Sciences (MTA KFKI RMKI), H-1525 Budapest 114, POBox 49, Budapest, Hungary}
\newcommand{\korea}{Korea University, Seoul, 136-701, Korea}
\newcommand{\kurchatov}{Russian Research Center ``Kurchatov Institute", Moscow, Russia}
\newcommand{\kyoto}{Kyoto University, Kyoto 606-8502, Japan}
\newcommand{\labllr}{Laboratoire Leprince-Ringuet, Ecole Polytechnique, CNRS-IN2P3, Route de Saclay, F-91128, Palaiseau, France}
\newcommand{\lawllnl}{Lawrence Livermore National Laboratory, Livermore, CA 94550, U.S.}
\newcommand{\losalamos}{Los Alamos National Laboratory, Los Alamos, NM 87545, U.S.}
\newcommand{\lpc}{LPC, Universit{\'e} Blaise Pascal, CNRS-IN2P3, Clermont-Fd, 63177 Aubiere Cedex, France}
\newcommand{\lund}{Department of Physics, Lund University, Box 118, SE-221 00 Lund, Sweden}
\newcommand{\muenster}{Institut f\"ur Kernphysik, University of Muenster, D-48149 Muenster, Germany}
\newcommand{\myongji}{Myongji University, Yongin, Kyonggido 449-728, Korea}
\newcommand{\nagasaki}{Nagasaki Institute of Applied Science, Nagasaki-shi, Nagasaki 851-0193, Japan}
\newcommand{\newmex}{University of New Mexico, Albuquerque, NM 87131, U.S. }
\newcommand{\nmsu}{New Mexico State University, Las Cruces, NM 88003, U.S.}
\newcommand{\ornl}{Oak Ridge National Laboratory, Oak Ridge, TN 37831, U.S.}
\newcommand{\orsay}{IPN-Orsay, Universite Paris Sud, CNRS-IN2P3, BP1, F-91406, Orsay, France}
\newcommand{\pnpi}{PNPI, Petersburg Nuclear Physics Institute, Gatchina, Leningrad region, 188300, Russia}
\newcommand{\riken}{RIKEN Nishina Center for Accelerator-Based Science, Wako, Saitama 351-0198, Japan}
\newcommand{\rikjrbrc}{RIKEN BNL Research Center, Brookhaven National Laboratory, Upton, NY 11973-5000, U.S.}
\newcommand{\rikkyo}{Physics Department, Rikkyo University, 3-34-1 Nishi-Ikebukuro, Toshima, Tokyo 171-8501, Japan}
\newcommand{\saispbstu}{Saint Petersburg State Polytechnic University, St. Petersburg, Russia}
\newcommand{\saopaulo}{Universidade de S{\~a}o Paulo, Instituto de F\'{\i}sica, Caixa Postal 66318, S{\~a}o Paulo CEP05315-970, Brazil}
\newcommand{\seoulnat}{System Electronics Laboratory, Seoul National University, Seoul, Korea}
\newcommand{\stonybrkc}{Chemistry Department, Stony Brook University, Stony Brook, SUNY, NY 11794-3400, U.S.}
\newcommand{\stonycrkp}{Department of Physics and Astronomy, Stony Brook University, SUNY, Stony Brook, NY 11794, U.S.}
\newcommand{\subatech}{SUBATECH (Ecole des Mines de Nantes, CNRS-IN2P3, Universit{\'e} de Nantes) BP 20722 - 44307, Nantes, France}
\newcommand{\tenn}{University of Tennessee, Knoxville, TN 37996, U.S.}
\newcommand{\titech}{Department of Physics, Tokyo Institute of Technology, Oh-okayama, Meguro, Tokyo 152-8551, Japan}
\newcommand{\tsukuba}{Institute of Physics, University of Tsukuba, Tsukuba, Ibaraki 305, Japan}
\newcommand{\vandy}{Vanderbilt University, Nashville, TN 37235, U.S.}
\newcommand{\waseda}{Waseda University, Advanced Research Institute for Science and Engineering, 17 Kikui-cho, Shinjuku-ku, Tokyo 162-0044, Japan}
\newcommand{\weizmann}{Weizmann Institute, Rehovot 76100, Israel}
\newcommand{\yonsei}{Yonsei University, IPAP, Seoul 120-749, Korea}
\affiliation{\abilene}
\affiliation{\banaras}
\affiliation{\bnlphys}
\affiliation{\caucr}
\affiliation{\cns}
\affiliation{\colorado}
\affiliation{\columbia}
\affiliation{\dapnia}
\affiliation{\debrecen}
\affiliation{\elte}
\affiliation{\fsu}
\affiliation{\gsu}
\affiliation{\hiroshima}
\affiliation{\ihepprot}
\affiliation{\illuiuc}
\affiliation{\isu}
\affiliation{\jinrdubna}
\affiliation{\kaeri}
\affiliation{\kek}
\affiliation{\kfki}
\affiliation{\korea}
\affiliation{\kurchatov}
\affiliation{\kyoto}
\affiliation{\labllr}
\affiliation{\lawllnl}
\affiliation{\losalamos}
\affiliation{\lpc}
\affiliation{\lund}
\affiliation{\muenster}
\affiliation{\myongji}
\affiliation{\nagasaki}
\affiliation{\newmex}
\affiliation{\nmsu}
\affiliation{\ornl}
\affiliation{\orsay}
\affiliation{\pnpi}
\affiliation{\riken}
\affiliation{\rikjrbrc}
\affiliation{\rikkyo}
\affiliation{\saispbstu}
\affiliation{\saopaulo}
\affiliation{\seoulnat}
\affiliation{\stonybrkc}
\affiliation{\stonycrkp}
\affiliation{\subatech}
\affiliation{\tenn}
\affiliation{\titech}
\affiliation{\tsukuba}
\affiliation{\vandy}
\affiliation{\waseda}
\affiliation{\weizmann}
\affiliation{\yonsei}
\author{S.~Afanasiev} \affiliation{\jinrdubna}
\author{C.~Aidala} \affiliation{\columbia}
\author{N.N.~Ajitanand} \affiliation{\stonybrkc}
\author{Y.~Akiba} \affiliation{\riken} \affiliation{\rikjrbrc}
\author{J.~Alexander} \affiliation{\stonybrkc}
\author{A.~Al-Jamel} \affiliation{\nmsu}
\author{K.~Aoki} \affiliation{\kyoto} \affiliation{\riken}
\author{L.~Aphecetche} \affiliation{\subatech}
\author{R.~Armendariz} \affiliation{\nmsu}
\author{S.H.~Aronson} \affiliation{\bnlphys}
% ...
\collaboration{PHENIX Collaboration} \noaffiliation

\date{\today}

\begin{abstract}

Bose-Einstein correlations of charged kaons are used to probe Au+Au 
collisions at $\sNN$ = 200 GeV and are compared to charged pion 
probes, which have a larger hadronic scattering cross section. Three 
dimensional Gaussian source radii are extracted, along with a 
one-dimensional kaon emission source function. The centrality 
dependences of the three Gaussian radii are well described by a 
single linear function of $\Npc$ with zero intercept. Imaging 
analysis shows a deviation from a Gaussian tail at $r\gtrsim10$ fm, 
although the bulk emission at lower radius is well-described by a 
Gaussian.  The presence of a non-Gaussian tail in the kaon source 
reaffirms that the particle emission region in a heavy ion collision 
is extended, and that similar measurements with pions are not solely 
due to the decay of long-lived resonances.

\end{abstract}

	
\maketitle

Experiments at the Relativistic Heavy Ion Collider (RHIC) at 
Brookhaven National Laboratory have revealed that collisions of Au 
ions at $\sNN$ = 200 GeV produce a new form of matter which is 
opaque to jets and exhibits anisotropic flow consistent with perfect 
fluid hydrodynamics~\cite{Adcox:2004mh,Hirano:2005wx}. Studies of 
the space-time evolution of the collisions are needed to elucidate 
the properties of the hot, dense, and strongly interacting matter, 
probe the time scale and degree of thermalization, and investigate 
the order of the deconfinement phase transition.

Two-particle interferometry, also known as HBT after the radio 
astronomers R. Hanbury Brown and R.Q. Twiss~\cite{Hanbury:1954wr}, 
is a powerful tool for measuring the space-time extent of 
particle-emitting sources. In elementary particle and nuclear 
physics, enhanced production of like-sign pions with small relative 
momenta was discovered experimentally and explained by the 
Bose-Einstein symmetrization of identical 
bosons~\cite{Goldhaber:1960sf}. Correlations are produced by the 
combination of quantum mechanical interference of identical 
particles and strong and/or electromagnetic final state interactions 
such as Coulomb repulsion for same-sign charged pairs. HBT radii 
refer to Gaussian measures of source sizes on the femtometer scale.

Although the traditional HBT analyses are constrained by the 
assumption of a Gaussian distribution of particle emission, recent 
detailed measurements of pion emission sources using an imaging 
technique show a non-Gaussian structure for the two-particle source 
region above $\sim$20 fm~\cite{Adler:2006as}, suggesting the 
possibility that decays of long-lived resonances or a temporal 
component of the source contribute to the non-Gaussian 
tail~\cite{Afanasiev:2007kk}.  While charged pions are strongly 
affected by rescattering among hadrons and decays of hadronic 
resonances, charged kaons have smaller rescattering cross sections 
than charged pions ($\sigma_{K-N}<\sigma_{\pi-N}$) and are less 
affected by resonance decays. 
Until recently, no full hydrodynamic calculation has accurately
predicted particle spectra and HBT radii for pions, and none can
simultaneously describe the momentum asymmetry measurements of flow.
The measurement of the kaon source describe herein add an important
new constraint to address this
``HBT puzzle''~\cite{Soff:2001hc,Bernard:1997bq}.

An angle-averaged one-dimensional Gaussian measurement of 
correlations of neutral kaons by STAR~\cite{Abelev:2006gu} suggests 
that the transverse mass dependence for neutral kaons and charged 
pions falls on one universal curve. In this paper, 3D Gaussian HBT 
correlations of like-sign kaons are presented in three transverse 
momentum bins 
for $0.3<\pT<1.5$ GeV/$c$ and three collision centrality bins. The 
resulting HBT radius parameters for kaons are compared to those of 
like-sign pion pairs~\cite{Adler:2004rq}. In addition we present 1D 
emission source functions for charged kaons in relativistic 
heavy-ion collisions.

This analysis of 2004 data from the PHENIX detector~\cite{Adcox:2003zm}
uses $\sim$~600 million minimum bias events, which are 
triggered by the coincidence of the Beam-Beam Counters (BBC) and 
Zero-Degree Calorimeters (ZDC) with collision vertex $|z|<30$~cm.
A Monte Carlo Glauber 
model~\cite{Glauber:1970jm} is used to match the observed BBC and 
ZDC distributions and to bin the data according to the number of 
nucleons participating in the collisions, $\Np$.

Charged kaons are tracked and identified using the drift chamber 
(DC), pad chambers (PC1,PC3) and PbSc Electromagnetic Calorimeters 
(EMCal) to cover pseudorapidity $|\eta|<0.35$ and azimuthal 
angle $\Delta\phi=\pi/2$ ($\Delta\phi=\pi/4$) in one (and the other)
central arm.  A track model provides 
a 3-dimensional trajectory and momentum vector for charged particles 
based on DC and PC1 information with a momentum resolution of 
$\delta p/p \simeq 0.7\% \oplus 1.0\%\times p$ (GeV/$c$). 
Backgrounds are reduced by requiring 2 $\sigma$ position match 
between track projections and EMCal hits, and 3 $\sigma$ match for 
PC3.  
Kaons are separated from pions up to $\pT \sim$0.9 GeV/$c$ using
timing information from BBC and EMC.
Particles at higher $\pT$ that fall within 2 $\sigma$ of the ideal
mass-squared for kaons but $\ge$ 3 $\sigma$ away from the peak for
pions or (anti-)protons are identified as kaons.
The contamination level is $\sim$4$\%$ from pions, and $\sim$1$\%$
from protons at $\pT\sim1.5$ GeV/$c$.

The correlation function is experimentally measured as 
$C_{2}(\bfq)=A(\bfq)/B(\bfq)$ where $A(\bfq)$ is the relative 
momentum ($\bfq$) distribution of actual pairs obtained by all 
possible combinations of pairs within the same events and $B(\bfq)$ 
is the background pair distribution from mixed events. Two-track 
detection inefficiencies for charged kaons that traverse the DC and 
EMCal in close proximity have been carefully studied with Monte-Carlo 
detector simulation and the actual pair distribution is corrected by 
the MC efficiency factors. After pair selection cuts to remove track 
splitting and merging (see~\cite{Adler:2004rq} for details), 
$\sim$15 million positive kaon pairs and 14 million negative kaon 
pairs remain.

\begin{figure}[thb]
\includegraphics[width=1.0\linewidth]{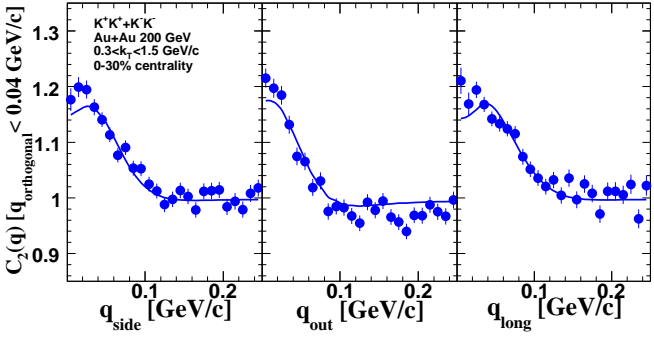}
\caption{\label{f:Kaon_3DC2} (color online).
3D correlation function of charged kaon pairs measured for
$0.3<\kT<1.5$ GeV/$c$ at $0-30\%$ centrality in Au+Au collisions at
$\sNN$ = 200 GeV. The curve is a 3D function is a Coulomb uncorrected 
fit with Eq. \ref{eq:Sinyukov} and sliced over the lowest 40 MeV/$c$ in 
the orthogonal directions.
}
\end{figure}

To measure multi-dimension source sizes, $\bfq$ is decomposed into 
standard ``side-out-long'' axes~\cite{Pratt:1984su}: for which $\ql$ 
is parallel to the beam axis, $\qo$ is parallel to the transverse 
momentum of the pair ($\kT=(p_{\rm 1T}+p_{\rm 2T})/2$), and $\qs$ is 
orthogonal to both $\ql$ and $\qo$. This analysis is performed in 
the Longitudinally Co-Moving System (LCMS) defined as $p_{\rm 
1Z}=-p_{\rm 2Z}$. For the treatment of charged kaons emitted away 
from the central region (core), we adopt an effective core-halo 
Coulomb correction, proposed by Bowler and 
Sinyukov~\cite{Sinyukov:1998fc}, in which the 3D Gaussian fit 
function is given by
\begin{eqnarray}
C_{2} = C^{core}_{2} + C^{halo}_{2} = \left[\lambda(1+G)\right]F_{\rm C} + \left[1-\lambda\right],
\label{eq:Sinyukov}
\end{eqnarray}
where the Coulomb correlation function $F_{\rm C}$ is iteratively 
evaluated from the Coulomb wave function of kaon pairs assuming a 
spherical Gaussian source.  The Gaussian correlation function in the 
side-out-long decomposition is determined by 
\begin{eqnarray}
G &=& {\rm exp}\left(-\RsS\qsS-\RoS\qoS-\RlS\qlS\right).
\label{eq:Gaussian}
\end{eqnarray}

The systematic error estimate incorporates a 
contribution from the Coulomb interaction of the source halo using a 
prescription developed by Maj and Mrowczynski~\cite{Maj:2007qw}. The 
fitted $\Rs$ and $\Rl$ are Gaussian measures of the spatial lengths 
of homogeneity, where particles of similar momenta are 
emitted~\cite{Makhlin:1987gm}, in the transverse and longitudinal 
directions at freeze-out. $\Ro$ contains a contribution from the 
duration of the particle emission in addition to the spatial 
length~\cite{Lisa:2005dd}. Note that an out-long cross-term vanishes 
in the expression for $G$ for our $|\eta|<0.35$ acceptance at 
midrapidity~\cite{Chapman:1994yv}. 
The fitted $\lambda$ is empirically defined and includes contributions
from mis-identified particles $(1-f)^2$ along with components of the
source that are not well resolved by the Gaussian fit.

Figure~\ref{f:Kaon_3DC2} shows the 3D correlation 
function of charged kaons without the Coulomb correction measured 
for $0.3<\kT<1.5$ GeV/$c$ at $0-30\%$ centrality in Au+Au collisions 
at $\sNN$ = 200 GeV and the fit with Eq.~(\ref{eq:Sinyukov}).  
Separate fits to the $2K^+$ and $2K^-$ correlation functions were 
performed, yielding consistent results for all $\kT$ and centrality 
bins.

\begin{figure}[thb]
\includegraphics[width=1.0\linewidth]{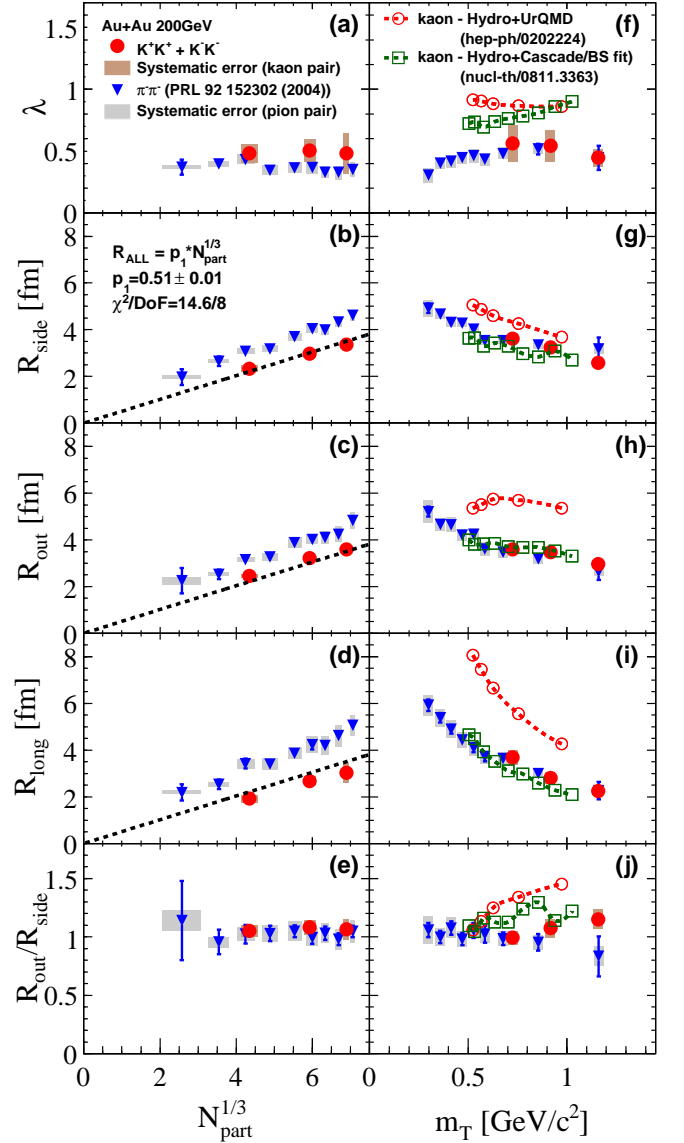}
\caption{\label{f:Kaon_3Dradii_mTNp} (color online).
3D Gaussian HBT radius parameters extracted from
Eq.(\ref{eq:Sinyukov}) for charged kaon pairs as a function
of (left) $\Npc$ measured for $0.3<\kT<1.5$ GeV/$c$ and 
(right) $\mT$ measured for $0-30\%$ centrality in Au+Au collisions at 
$\sNN$ = 200 GeV.  In panels (b)-(d) the dashed line is a fit with 
$p_{1}*\Npc$.
}
\end{figure}

Panels (a)-(e) in Fig.~\ref{f:Kaon_3Dradii_mTNp} show the HBT radius 
parameters of charged kaons for $0.3<\kT<1.5$ GeV/$c$ as functions of 
$\Npc$, which is proportional to the transverse radius of the initial 
collision volume. Similar to pions, kaon radii are well described by 
linear functions of $\Npc$.  Because initial fits yielded slopes that 
were consistent for all radii and intercepts that were consistent with 
zero, the all three radii were fit to a single linear function with 
zero intercept: $R_{i} = p_1* \Npc$, with $p_1=0.51 \pm 0.01$ and 
$\chi^2/{\rm ndf} = 14.6/8$.  We note that similar fits to the pion 
radii yield non-zero intercepts. Although pions and kaons are measured 
in a similar $\kT$ range ($0.2<\kT<2.0$ GeV/$c$ for pions), the higher 
transverse mass ($\mT=\sqrt{\kT^{2}+m^{2}}$) for kaons ($\mmT 
\sim$0.89 GeV/$c^2$) than for pions ($\mmT \sim$0.47 GeV/$c^2$) leads 
to smaller radii, as expected from $\mmT$ 
scaling~\cite{Makhlin:1987gm,Akkelin:1995gh,Bearden:1996dd,Csanad:2008gt}.

Panels (f)-(j) in Fig.~\ref{f:Kaon_3Dradii_mTNp} show the $\mT$ 
dependence of the radius parameters for kaons in 3 different $\mT$ 
bins at $0-30\%$ centrality, compared with pions~\cite{Adler:2004rq}. 
HBT radii are quite consistent at the same $\mT$, clearly indicating 
that the radii follow $\mT$ scaling.  The $\Ros$ ratio for kaons is 
$\sim$1.0-1.2 which is consistent with the value for pions at low 
$\mT$. Kaon HBT results from a 2D+1 hybrid, hydrodynamic + UrQMD 
calculation (open circles)~\cite{Soff:2002qw} show slightly larger 
sidewards radii than the data, and the outwards and longitudinal 
components are too large by a factor of 2-3. A more recent 1D+1 hybrid 
calculation~\cite{Pratt:2008qv} for kaons (open squares), which 
assumes flat rapidity distribution and axial symmetry, compares more 
favorably, matching all radii to within systematic and statical 
errors. This calculation incorporates pre-equilibrium flow and a 
lattice-inspired equation of state, which are two features lacking in 
earlier calculations of HBT radii. Although promising, these 
theoretical results remain to be verified with full 3D+1 calculations 
that can also reproduce the elliptic flow.

Recent femtoscopic measurements~\cite{Adler:2006as,Afanasiev:2007kk}, 
which use an 
imaging technique~\cite{Brown:1997ku} revealed that the emission 
source function of charged pions has a non-Gaussian tail which 
cannot be resolved with traditional Gaussian fitting techniques. In 
the imaging scheme, the correlation function is expressed by the 
Koonin-Pratt equation~\cite{Koonin:1977fh,Pratt:1990zq}
\begin{eqnarray}
\CqB - 1 = \int d\bf r \KernB\SrB,
\label{eq:Koonin-Pratt}
\end{eqnarray}
where the kernel $\KernB$ is the relative wave function
as $|\Phi^{(-)}_{\bf q}(\bf r)|^{2}-1$ that describes the propagation
of pairs emitted with relative separation $\bf r$ and relative
momentum $\bf q$ in the Pair Center-of-Mass System (PCMS).
$\SrB$ is the emission source function of pairs.

\begin{figure}[thb]
\includegraphics[width=1.0\linewidth]{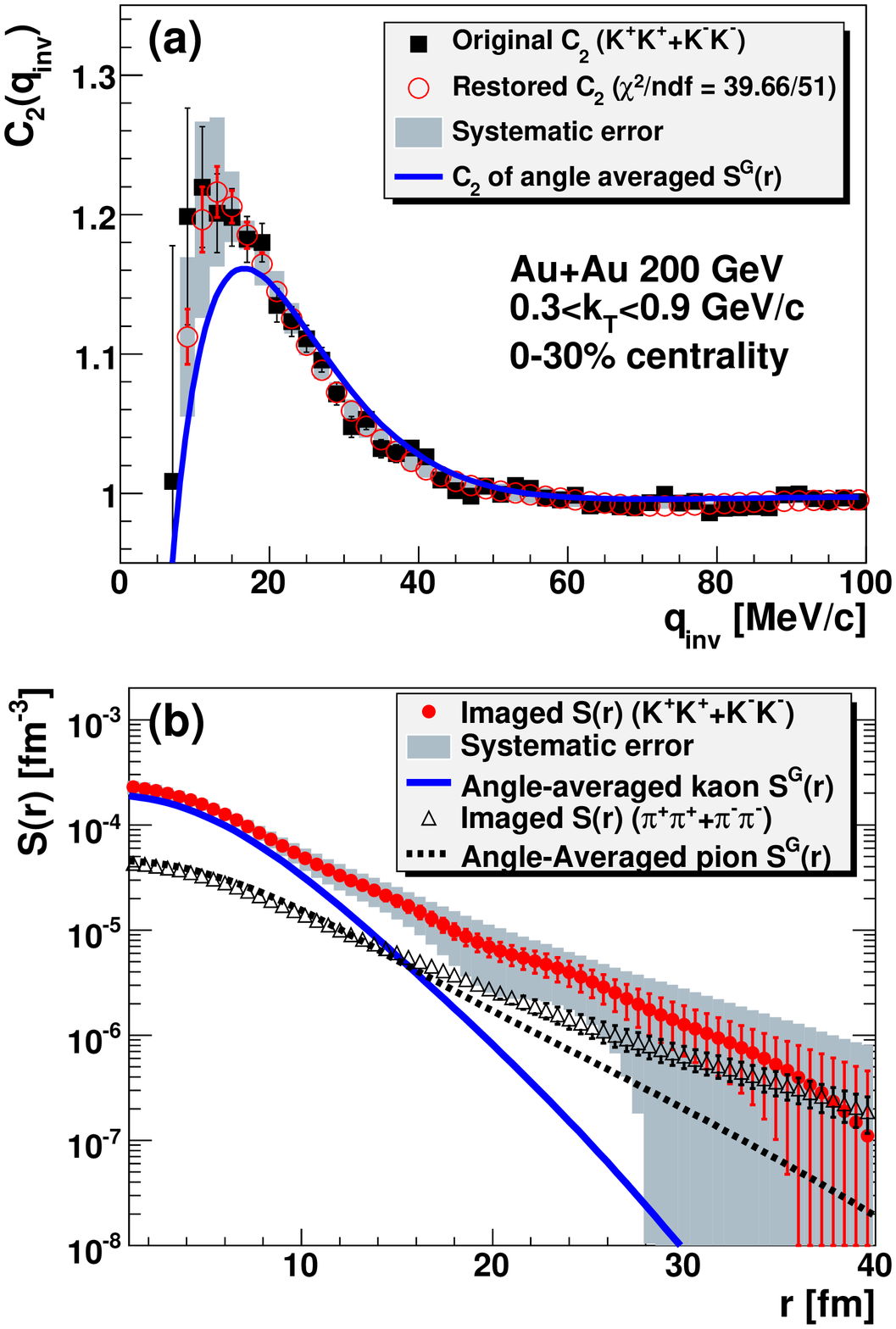}
\caption{\label{f:Kaon_1Dimaging} (color online).
(a) (filled squares) measured $\CqI$. (open circles) restored
$\CqI$ from imaged $\SrI$, compared with (solid curve)
angle-averaged Gaussian $\CqI$ for charged kaon
pairs measured for $0.3<\kT<0.9$ GeV/$c$ at $0-30\%$ central 
Au+Au collisions at $\sNN$ = 200 GeV.
(b) (filled circle) Imaged kaon $\SrI$ compared with
(solid curve) angle-averaged Gaussian $\SrI$.
Error bars are statistical only and boxes indicate the total
systematic errors. (open triangle) $\SrI$ for charged pion pairs 
for the same $\kT$ region.
For the pion $\SrI$, error bars include both statistical and
systematic errors.}
\end{figure}

The filled squares in Fig. \ref{f:Kaon_1Dimaging} (a) show the 1D 
kaon correlation as a function of the invariant relative momentum of 
the pair ($\qi=\sqrt{-(p_{1}-p_{2})^2}/2$). The 1D source function 
$\SrI$ imaged from $\CqI$ is shown by filled circles in Fig. 
\ref{f:Kaon_1Dimaging} (b).

In this analysis, input parameters that govern the imaging 
procedure~\cite{Brown:1997ku} were selected to minimize the \cisq 
between the data and the restored $\CqI(\cisq/{\rm ndf}\sim 1 $), 
shown by open circles in Fig. \ref{f:Kaon_1Dimaging} (a). The solid 
curve shows the traditional Gaussian source function, obtained by 
angle-averaging the 3D HBT radius parameters ($\lambda$, $\Rs$, 
$\Ro$, $\Rl$) in the PCMS frame, the same frame in which the imaging 
is performed.

The imaged $\SrI$ exhibits a non-Gaussian tail at
$r\gtrsim10$ fm.
This excess corresponds to the deficit in the $\qi\lesssim$ 20 MeV/$c$
region of the angle-averaged Gaussian curve of
Fig.~\ref{f:Kaon_1Dimaging} (a), and is also visible in the 3D
Gaussian slices in Fig.~\ref{f:Kaon_3DC2}.
The $\SrI$ for pions in the 
same $\kT$ range shows a similar trend. 
The deviation from a Gaussian in the shape of the $\SrI$ indicates
that the particle emission region is extended, and a similar
non-Gaussian tail in the pion source is not solely the result of long
lived resonance decays such as the $\omega$,  although a less
prominent contribution from the  $K^*$ is likely.
The observation of a more substantial non-Gaussian
tail for kaons than for pions is qualitatively consistent with a hadronic
resonance cascade model with a time dependent density
for an expanding source, in which the larger mean free path
for kaons leads to an extended emission
region~\cite{Csorgo:2005it,Csanad:2007fr}.
Detailed measurements with 3D HBT imaging of kaons, or 1D imaging of
more species probing different hadronic cross sections will determine
contributions from other kinetic effects to $\SrI$.


In summary, we have measured Bose-Einstein correlation functions of 
charged kaon pairs in Au+Au collisions at $\sNN$ = 200 GeV. The 3D HBT 
radii $\Rs$ and $\Rl$ are consistent for pions and kaons at the same $\Np$ 
and $\mT$. The 1D emission source function for kaons extracted by imaging 
shows a non-Gaussian tail at distances greater than 10 fm. This tail 
represents a direct measurement of the 1D length of homogeneity of the 
particle emission source and is not due primarily to resonance decays.


\begin{acknowledgments}


We thank the staff of the Collider-Accelerator and 
Physics Departments at BNL for their vital contributions.  
We acknowledge support from the
Office of Nuclear Physics in DOE Office of Science and NSF (U.S.A.), 
MEXT and JSPS (Japan), 
CNPq and FAPESP (Brazil), 
NSFC (China), 
IN2P3/CNRS, and CEA (France), 
BMBF, DAAD, and AvH (Germany), 
OTKA (Hungary), 
DAE (India), 
ISF (Israel), 
KRF and KOSEF (Korea), 
MES, RAS, and FAAE (Russia),
VR and KAW (Sweden), 
U.S. CRDF for the FSU, 
US-Hungarian NSF-OTKA-MTA, 
and US-Israel BSF.

\end{acknowledgments}



\end{document}